# Inhomogeneity of charge density wave order and quenched disorder in a high $T_c$ superconductor

G. Campi[1,2], A. Bianconi[2,1], N. Poccia[3,2], G. Bianconi[4], L. Barba[5], G. Arrighetti[5], D. Innocenti[6,2], J. Karpinski[6,7], N. D. Zhigadlo[7], S. M. Kazakov[7,8], M. Burghammer[9,10], M. v. Zimmermann[11], M. Sprung[11], A. Ricci[11,2]

1. Institute of Crystallography, CNR, via Salaria Km 29.300, Monterotondo Roma, I-00015, Italy.
2. Rome International Center for Materials Science, Superstripes, RICMASS, via dei Sabelli 119A, I-00185 Roma, Italy
3. MESA+ Institute for Nanotechnology, University of Twente, PO Box 217, 7500AE Enschede, Netherlands.
4. School of Mathematics, Queen Mary University of London, London E1, 4SN, UK.
5. Institute of Crystallography, Sincrotrone Elettra UOS Trieste, Strada Statale 14 - Km 163,5 Area Science Park, 34149 Basovizza, Trieste, Italy.
6. EPFL, Institute of Condensed Matter Physics, Lausanne CH-1015, Switzerland.
7. ETH, Swiss Federal Institute of Technology Zurich Laboratory for Solid state Physics CH-8093 Zurich CH
8. Department of Chemistry, M.V. Lomonosov Moscow State University, Moscow 119991, Russia
9. European Synchrotron Radiation Facility, B. P. 220, F-38043 Grenoble Cedex, France
10. Department of Analytical Chemistry, Ghent University, Krijgslaan 281, S12 B-9000 Ghent, Belgium
11. Deutsches Elektronen-Synchrotron DESY, Notkestraße 85, D-22607 Hamburg, Germany.

**It has recently been established that the high temperature (high-Tc) superconducting state coexists with short-range charge-density-wave order[1-11] and quenched disorder[12,13] arising from dopants and strain[14-17]. This complex, multiscale phase separation[18-21] invites the development of theories of high temperature superconductivity that include complexity[22-25]. The nature of the spatial interplay between charge and dopant order that provides a basis for nanoscale phase separation[19-21] remains a key open question, because experiments have yet to probe the unknown spatial distribution at both the nanoscale and mescoscale (between atomic and macroscopic scale). Here we report micro X-ray diffraction imaging of the spatial distribution of both the charge-density-wave 'puddles' (domains with only a few wavelengths) and quenched disorder in $HgBa_2CuO_{4+y}$, the single layer cuprate with the highest $T_c$, 95 kelvin[26-28]. We found that the charge-density-wave puddles, like the steam bubbles in boiling water, have a fat-tailed size distribution that is typical of self-organization near a critical point[19]. However, the quenched disorder, which arises from oxygen interstitials, has a distribution that is contrary to the usual assumed random, uncorrelated distribution[12, 13]. The interstitials-oxygen-rich domains are spatially anti-correlated with the charge-density-wave domains, leading to a complex emergent geometry of the spatial landscape for superconductivity.**

Although it is known that the incommensurate charge-density-wave (CDW) order in cuprates (copper oxides) is made of ordered, stripy, nanoscale puddles with an average of only 3-4 oscillations, information about the size distribution and spatial organization of these puddles has so far not been available. We present experiments that demonstrate that CDW puddles, have a complex spatial distribution and coexist with, but are spatially anticorrelated to, quenched disorder in $HgBa_2CuO_{4+y}$ (Hg1201). The sample we studied is a layered perovskite at optimum doping with oxygen interstitials y=0.12, tetragonal symmetry P4/mmm and a low misfit strain [14-16]. The X-ray diffraction (XRD) measurements (see Methods) show diffuse CDW satellites (secondary peaks surrounding a main peak) at $q_{CDW}$=(0.23a*, 0.16c*), in the b*=0 plane and $q_{CDW}$=(0.23b*, 0.16c*) in the a*=0 plane (where a*, b*, and c* are the reciprocal lattice units) around specific Bragg peaks, such as (1 0 8), below the onset temperature $T_{CDW}$=240 K (see Fig. 1a). The component of the momentum transfer $q_{CDW}$ in the $CuO_2$ plane (0.23a*) in this case is smaller than it is in the underdoped case (0.28a*)[5]. The temperature evolution of CDW-peak profile along a* (in the h direction; Fig. 1b) shows a smeared, glassy-like evolution below $T_{CDW}$. The CDW-peak intensity reaches a maximum at T=100 K, followed by a drop associated with the onset of superconductivity at T=$T_c$. We investigated the isotropic character of the CDW, in the *a-b* plane using azimuthal scans, as shown in Fig. 1c. We observed an equal probability of vertical and horizontally striped CDW puddles.

Our main result is the discovery of the statistical spatial distribution of the CDW-puddle size and density throughout the sample, which shows an emergent complex network geometry for the superconducting phase. We performed scanning micro X-ray diffraction (SµXRD) measurements (see Methods) to extend the imaging of spatial inhomogeneity previously obtained by scanning tunneling microscopy[7-9], from the surface to the bulk of the sample and from nanoscale to mesoscale spatial inhomogeneity. Clear evidence of the inhomogeneous spatial distribution of the CDW is provided by the observation of very different CDW-peak profiles collected at different illuminated sample spots (see Fig. 1d) corresponding to spots with "*large*" and "*small*" puddles.

We investigated the temperature dependence of CDW domains by recording the CDW-peak intensity and its full-width at half maximum (FWHM) during cooling from 280 K to 85 K. We collected the data in two different places on the sample corresponding to "*large*" and "*small*" CDW puddles. Fig. 1e, f shows the temperature evolution of population (intensity), the number of oscillations $h_{CDW}/\Delta h_{CDW}$ (where $h_{CDW}$ and $\Delta h_{CDW}$ are the position and the FWHM of the CDW peak profile in units of a*, respectively) and in plane puddles size $\xi_a$ (along th a axis) in *large* (red filled circles) and *small* (black filled squares) CDW puddles.

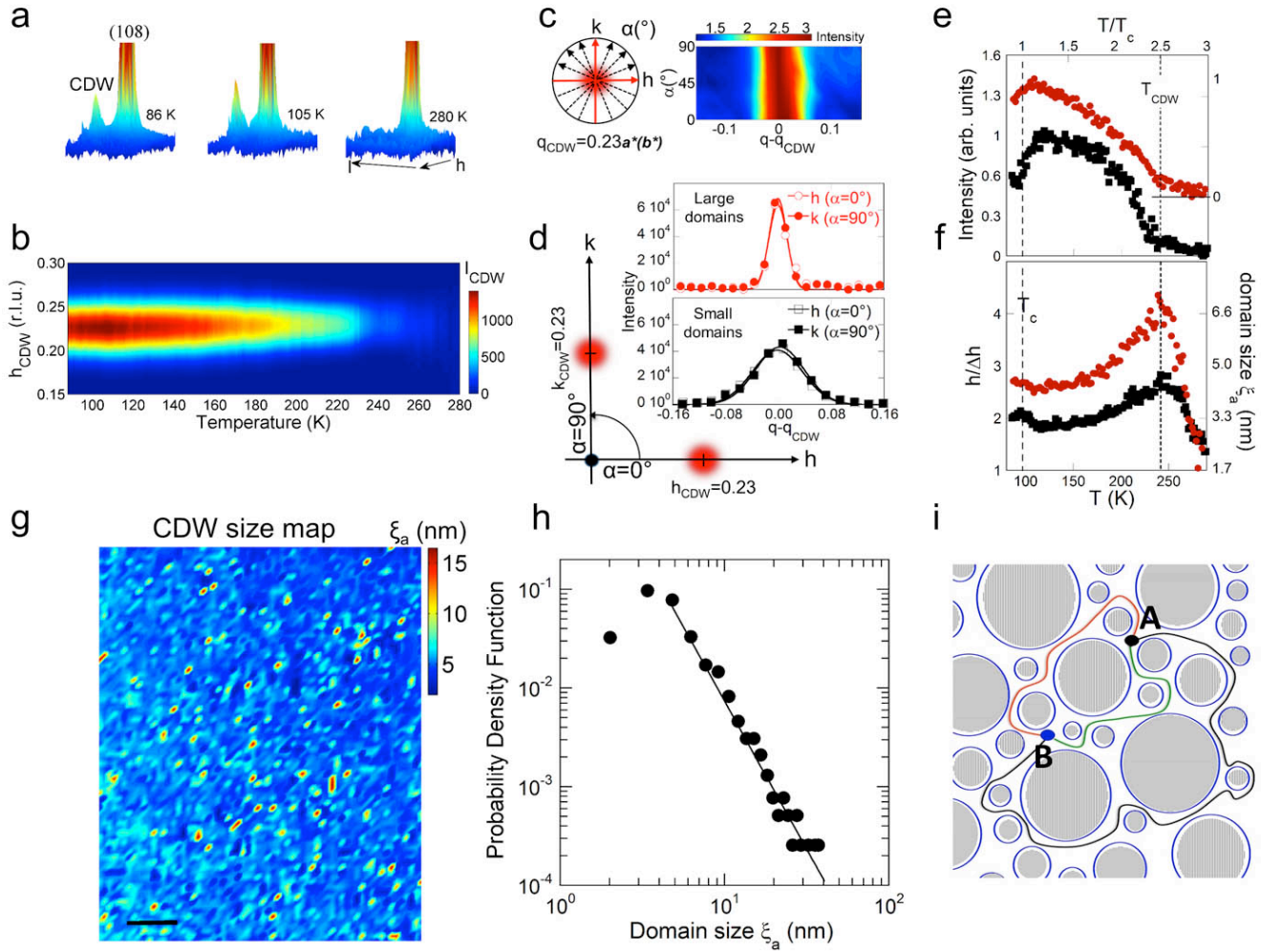

**Figure 1| Temperature dependence and spatial distribution of CDW puddles in Hg1201. a** The CDW satellite near the (108) Bragg peak appears below 240 K. **b** Temperature dependence of CDW- peak profiles along h. The CDW-peak Intensity $I_{CDW}$ is measured as the number of counts subtracted by the background. **c** The $q_{CDW}$ = (0.23a*(b*), 0.17c*) peak profile at different azimuthal angles α showing the peak isotropy. **d** Two typical CDW peaks collected at two different places in the same crystal. Red solid circles correspond to the diffraction profile from an illuminated part of the sample with large CDW puddles (red in g); black filled squares correspond to an illuminated part of the sample with small CDW puddles; (blue in g); the solid lines are Gaussian fits. **e** The CDW-peak intensity as a function of temperature, at the two different places on the sample corresponding to large (red filled circles, right axis) and small (black filled squares, left axis) CDW puddles. The dashed line corresponds to T = $T_c$ and the dotted line to T = $T_{CDW}$. **f** Evolution of the number of CDW oscillations ($h_{CDW}/\Delta h_{CDW}$) inside a CDW puddle and the CDW domain size along the a axis, $\xi_a$. **g, h** Spatial map (**g**) and probability Density Function (PDF) (**h**) of the CDW-puddles size. Scale bar in (g) 10 μm. **i** A schematic of non-equivalent paths, running in the interface space between CDW puddles, connecting point A to point B in the emergent complex hyperbolic geometry[29,30], for the superconducting current.

The broad phase transition appears to be arrested, as indicated by the size of the CDW puddles $\xi_a=1/\Delta h_{CDW}$, which does not diverge below $T_{CDW}$. This behavior is typical of low dimensional systems with quenched disorder.

A map representing the spatial organization of the CDW-puddle size is shown in Fig. 1g. The probability density function (PDF) of the in-plane CDW-puddle size $\xi_a$ is shown in Fig. 1h. The PDF has a long fat tail that extends over an order of magnitude and is fitted by $PDF(\xi_a) \approx \xi_a^{-\alpha(CDW)} \exp(-\xi_a/\xi_t)$ where $\alpha_{CDW}=2.8\pm0.1$ is the critical exponent of the puddle-size power law distribution and $\xi_t > 40$ nm. Although we can determine that the average size of CDW puddles is 4.3 nm (in agreement with previous works), $PDF(\xi_a)$ has a non-Gaussian shape and rare, larger puddles reaching sizes of 40 nm are detected. Our finding of a fat-tailed distribution for the CDW-puddle size is in agreement with previous results in the nanoscale obtained by STM[9]. Such structures, where spontaneous breaking of both translational symmetry (CDW electronic crystalline phase) and gauge symmetry (superconductivity) coexist, have been called superstripes[19]. The distribution of the CDW puddles we have found introduces a substantial topological change to the available space for superconductivity: the current running from a point A to a point B of the material can take different paths (see Fig. 1i) that are not topologically equivalent [29] thus forming an emergent hyperbolic geometry[30].

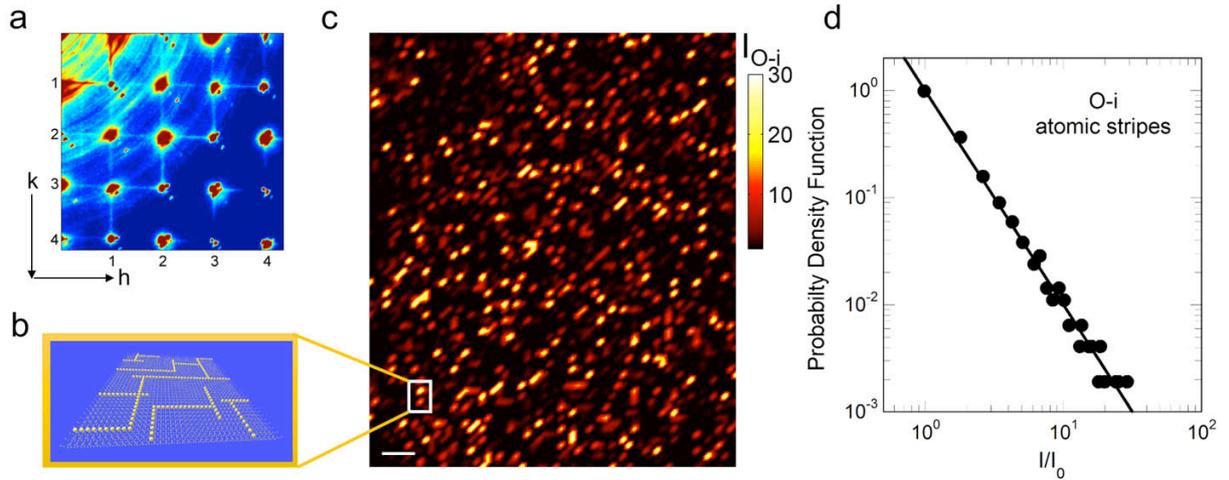

**Figure 2| Correlated quenched disorder due to $O_i$ atomic stripes in Hg1201. a** A portion of the h-k diffraction pattern. Resolution-limited streaks connect the Bragg peaks, owing to the formation of $O_i$ stripes in the $HgO_y$ spacer layers. **b** Schematic representation of the atomic $O_i$ stripes. **c** SµXRD map of the region of (**a**) showing the relative $O_i$ streak intensity $I_{Oi}$. The bright (dark) spots correspond to sample regions with high (low) density of $O_i$ atomic stripes, called $O_{Oi}$ rich (poor) regions. The bar corresponds to 5 µm. (d) Probability Density Function (PDF) calculated from the $O_i$ streaks intensity map.

To investigate the interplay between the CDW puddles and the quenched disorder, we studied the spatial distribution of oxygen defects. The quenched lattice disorder is due to oxygen interstitials ($O_i$), which form $O_i$ atomic stripes in the $HgO_y$ layers, in agreement with previous experiments[27, 28]. $HgBa_2CuO_{4+y}$ (ref. 28), like $YBa_2Cu_3O_{6+y}$ (ref. 15) and $La_2CuO_{4+y}$ (ref. 14, 16) shows $T_c$ variations, owing to the effect of the spatial organization of $O_i$ on superconductivity. The average $O_i$ self-organization was detected by high-energy XRD (see Methods). Figure 2a shows the (0<h<5, 0<k<4) portion of reciprocal space, where there is strong evidence of diffuse streaks running along the a* and b* directions and crossing all the Bragg peaks. Our high energy XRD data confirm the formation of $O_i$ stripes intercalated between the $CuO_2$ planes, both in the (100) and (010) directions[26].

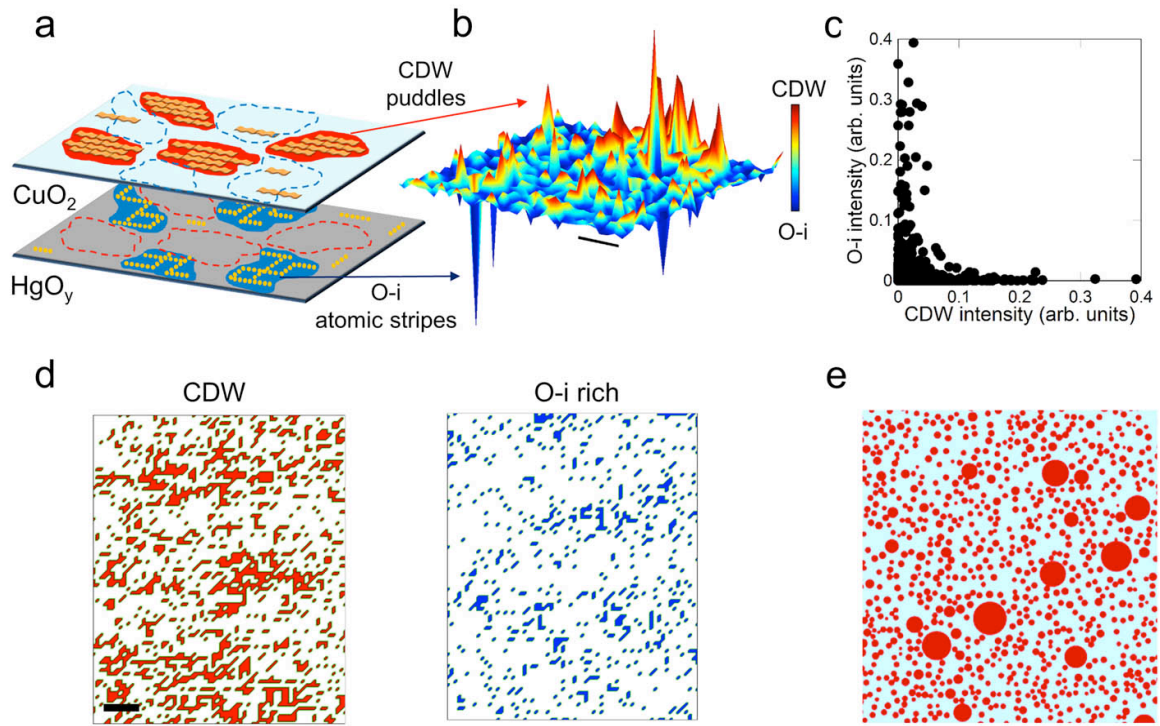

**Figure 3| Spatial anti-correlation between CDW-rich and the $O_i$-rich regions. a** The CDW-rich regions (red) on the $CuO_2$ planes and $O_i$-rich regions (blue) on the $HgO_y$ layers. **b** Surface plot of the difference map (see methods) between the CDW peak and $O_i$ streaks intensity. The positive (green to-red) values indicate the CDW-rich regions, while the negative (green to blue) values correspond to $O_i$-rich regions. Scale bar, 5 μm. **c** Scatter plot of the $O_i$ versus CDW intensity demonstrating the negative correlation between CDW-puddles and $O_i$ stripes populations. **d** Segmentation of the difference map in (**b**) highlighting the network of CDW-rich domains (left panel) and $O_i$-rich regions (right panel). Scale bar, 10 μm. **e** A schematic of the nanoscale texture formed by CDW-rich regions (red spots) and the "charge $O_i$-rich" (light-blue area) which define an interface space and loci of the superconductivity with a complex non-Euclidean geometry[29,30].

The spatial distribution of the intensity of the streaks was obtained by SμXRD (see Methods). We measured the reciprocal a*-c* plane (or b*-c* plane) around the (006) Bragg peak in reflection geometry. The $O_i$ stripes in Hg1201 run along the a*(b*) direction with no correlation along the c* direction; therefore, they also lead to streaks also on the a*c* plane. A schematic of

$O_i$ atomic stripes is shown in Fig. 2b. In Fig. 2c we show the spatial map of the streak intensity. The picture shows rich (bright-yellow) and poor (dark-black) regions of $O_i$ stripes. The PDF of $O_i$-rich regions in Fig. 2d can be fitted by $PDF(I/I_0) \approx I/I_0^{-\alpha(O_i)} \exp(I/I_t)$ where $I_0$ is the average intensity value, $\alpha(O_i)=2.0 \pm 0.1$ is the critical exponent and $I_t > 20$.

In Fig. 3 we present results on the spatial interplay between CDW-rich regions and $O_i$-rich regions. We calculated the 'difference map' (see Methods) between CDW peaks and $O_i$ diffuse streaks. The poor CDW regions on the $CuO_2$ basal plane correspond to $O_i$-rich regions on the $HgO_y$ layers, as illustrated in Fig. 3a. The CDW puddles and $O_i$ rich regions give rise to the positive and negative peaks, respectively, in the surface plot shown in Fig. 3b. The spatial anti-correlation is evident from the scatter plot of $O_i$ intensity versus CDW intensity (Fig. 3c). As $O_i$ intensity increases, the CDW intensity decreases, and vice versa. This is consistent with the fact that excess $O_i$ means higher doping and high doping does not favour stripy, underdoped short-range CDW order. Fig. 3d shows the two maps obtained through the segmentation of the difference map, and provides direct image of how doping poor (CDW-rich regions are shown in red) and doping-rich ($O_i$-rich regions are shown in blue) phases are arranged in different regions of the material. Figure 3e illustrates the nanoscale configuration of CDW-puddles (red spots) in the $CuO_2$ plane using the experimental distribution of CDW size; this distribution generates 'holes' in the space available for the free electrons (light blue area). This space is topologically interesting: there are an infinite number of ways for a current path to connect a point A to a point B around the CDW puddles, which are not only distinguished by the number of times a path goes around a single hole, but also by the way the path passes through the pattern of CDW puddles[27,28]. The complex space that emerges from the mesoscopic phase separation both in the spacer layers and in the $CuO_2$ plane, substantially changes (1) the dielectric constat that controls the long-range Coulomb interaction that is relevant for phase separation near a Lifshitz transition[19], (2) the dielectric constant that is relevant to electron-electron interaction in the pairing and (3) the geometrical and topological properties of the space that is available for the overall phase coherence of the macroscopic quantum condensate that is made up of multiple condensates at the nanoscale with a single critical temperature[19,23].

The work offers new insight into the complexity of nanoscale phase-separation phenomena in high temperature superconductors. More generally, our results deal with the effects of quenched disorder in phase transitions. A phase transition that would be of first order in the clean limit gets smeared into a continuous-looking transition in the presence of a random, Gaussian distributed, quenched disordered background[12,13]. Here the disorder itself is not randomly distributed, but has long-tailed probability density function, leading to correlated disorder. Even in the "ideal" single layer cuprate superconductor $HgBa_2CuO_{4+y}$ at optimum doping ($T_c$=95K), the CDW self-organizes in puddles, forming the inhomogeneous landscape with an emergent complex network geometry. Our results provide further evidence for the universality of mesoscale phase separation even in the most optimized superconducting copper oxides, which implies that the superconductivity will be non-uniform throughout what is a granular medium.

## Acknowledgements

We acknowledge ESRF, ELETTRA and DESY synchrotron facilities for radiation beamtime and support. We thank beamline scientists for extraordinary experimental help. We acknowledge Calypso program for traveling support. We acknowledge support of the Superstripes Institute. N.P. acknowledges for financial support the Marie Curie IEF project for career development.

## Author Contributions

All authors have contributed to essential portions of this work; A.B., G.C. & A.R. equally contributed to this work. The experiment was conceived by A.B., G.C., N.P., G.B. & A.R. The Hg1201 crystals were grown at ETH by S.M.K., J.K.; N.D.Z. and N.P. performed magnetic characterization of Hg1201 single crystals; experiments at DESY were performed by A.R., G.C., N.P., M.Z, D.I.; experiments at ESRF were performed by A.R., N.P., G.C., A.B., M.B.; experiments at Elettra were performed by G.C., L.B, G.A., A.B. & A.R., the data analysis has been carried out by G.C., A.B., G.B. & A.R. All authors discussed the results and contributed to the writing of the manuscript.

# Methods

**Sample preparation and characterization.** The $HgBa_2CuO_{4+y}$ (Hg1201) crystal with y=0.12, grown at ETH[24], shows a sharp superconducting transition at $T_c$=95 K. Crystal structure has P4/mmm symmetry with lattice parameters a=b=0.38748(5) nm and c=0.95078(2) nm at T=100K.

**X-Ray Diffraction measurements at the XRD1 beamline, ELETTRA.** In order to identify the CDW order in Hg1201 single crystal we have used X-ray diffraction at the XRD1 beam-line at the ELETTRA synchrotron radiation facility in Trieste, Italy, tuning the photon energy between 13 KeV and 16 KeV with a 200 x 200 $\mu m^2$ beam size. Only selected reflections show clear CDW satellites, in agreement with ref. 2. We focused on the CDW satellite located at $q_{CDW}$=(0.23,0,0.16) around the (1,0,8) Bragg reflection which appeared as the sample was cooled below 240K. Typical diffraction pattern collected at 85 K, 105 K and 280 K, are shown in Fig. 1a. To get a direct view of the temperature dependence of the CDW satellite reflection for T=280-85K, a two dimensional color-plot of the CDW peak profile along the **a*** direction as a function of temperature, is shown in Fig. 1b.

**High Energy X-Ray Diffraction measurements at the BW5 beamline.** High energy X-Ray Diffraction measurements were collected using transmission geometry and X-ray energy of 100 KeV. A single SiGe(111) gradient monochromator was used. The beam size was 200 x 200 $\mu m$. We used a vertical rotation axis and the single crystal c-axis has been oriented parallel to the direction of the incoming X-ray beam. The diffraction patterns were collected by an area detector in the temperature range of 20-300K. In this geometry we can probe the lattice fluctuations on the *a-b* plane. The possible CDW peak anisotropy was seen by azimuthal scans (0°<α<90°) as shown in the schematic in Fig. 1c. The CDW peak amplitude and FWHM do not change substantially as a function of α as shown in the color plot of the diffraction profile (as a function of α) in Fig. 1(c), instead, this plot shows the CDW planar isotropy in agreement with the tetragonal P4/mmm symmetry of the lattice. The presence of resolution-limited streaks, connecting the Bragg peaks, owing to the organization of single $O_i$ stripes in the mercury spacer layer are shown in Figure 2a. This figure shows a portion of the h-k diffraction pattern that was collected at DESY. The spatial distribution of the $O_i$ atomic stripes in the $HgO_y$ spacer layers which cause one dimensional doping and lattice spatial inhomogeneity, does not vary with temperature below 250 K leading to quenched disorder at the onset of charge-density phase.

**Scanning micro X-ray diffraction measurements at the ID13 beamline.** Scanning micro X-ray diffraction experiments were performed in reflection geometry using the ID13 beamline at ESRF, Grenoble, France. We applied incident X-ray energy of 13 KeV. By moving the sample under a 1 μm$^2$ focused beam with an x-y translator, we scanned a sample area of 65 x 80 μm$^2$ collecting 5200 different diffraction patterns at T=100 K. For each scanned point of the sample the **q$_{CDW}$** peak profile was extracted; the FWHM along **a\*(b\*)** and **c\*** direction were evaluated to obtain the domain size of the charge ordered regions along the **a(b)** and **c** crystallographic axes. Two quite different CDW peak profiles in the same crystal measured along the **a\*(b\*)** direction in the *a\*-c\*(b\*-c\*)* plane are shown in Fig.1c. Here we show two typical profiles collected at two different spatial locations in the same crystal corresponding to large (red circles) and small (black squares) puddles. The continuous lines are the Gaussian fits to the data. The different amplitude and FWHM (0.033±0.001 r.l.u. and 0.089±0.001 r.l.u. in the upper and lower panel of Fig. 1d, respectively) of the two peaks, which correspond to large and small CDW puddles, provide evidence of a strong inhomogeneity in the CDW spatial distribution. The peak profiles do appear the same along the **a\*** and **b\*** direction, confirming the peak isotropy in the basal plane of the tetragonal lattice. The intensity of the CDW satellite as a function of temperature, measured at two different locations on the sample corresponding to CDW large (red) and small (black) puddles are shown in Fig. 1e. The vertical lines represent the superconducting temperature T$_c$ and the CDW onset temperature T$_{CDW}$. The order-disorder transition is very broad, which indicates the role of the quenched disorder owing to the presence of defects. Moreover, the CDW intensity shows a clear drop, around T$_c$ that appears to depend on the CDW puddle size. The temperature dependence of the number of CDW oscillations, h$_{CDW}$/Δh$_{CDW}$, inside a single puddle and the domain size of a single puddle along the *a(b)* axis (ξ$_a$) are shown in Fig. 1f. (Δh$_{CDW}$ and h$_{CDW}$ are the FWHM and the location along *a\** of the CDW peak; the domain size along the *a*-axis (*b*-axis) is given by the correlation length ξ$_a$. The inhomogeneity of the CDW distribution is depicted by the 65 x 80 μm$^2$ XRD map of the (nanoscale) size of CDW domains in Fig. 1g. This figure shows loci of large (red-yellow area), and small (blue area) CDW puddles. The scale bar corresponds to 10 μm. Using the ID13 microfocus beamline at ESRF we can also detect the spatial distribution of the quenched disorder. Fig 2c shows the SμXRD map of integrated intensity of the streaks of O$_i$ stripes. The bright (dark) spots correspond to sample regions with high (low) density of O$_i$ atomic stripes, called O$_i$ rich (poor) regions. The scale bar

1s 10 μm. Fig. 2d shows the Probability Density Function (PDF) of the $O_i$ streaks intensity that was obtained from the μXRD map. This plot shows the probability distribution for the $O_i$ rich regions. The experimental set-up allows us to investigate the spatial interplay between CDW puddles in the $CuO_2$ plane and the $O_i$ rich domains in the $HgO_y$ layers shown in Fig. 3. We measured the 'difference map' ($I_{CDW}$-$I_{Oi}$) where $I_{CDW}$ and $I_{Oi}$ are the intensity of $q_{CDW}$ peak and the $O_i$ diffuse streaks, respectively, normalized to [0 1]. The surface plot of this difference map is shown in Fig. 3b. The positive (green to-red) peaks indicate CDW-puddles-rich regions, and the negative (green to blue) peaks indicate $O_i$ rich regions. The spatial anti-correlation between CDW puddles and $O_i$ atomic stripes is obtained by segmentation of the difference map. We use this segmentation to visualize the phase separation owing to the network of CDW-rich domains, which correspond to "charge poor" domains in the $CuO_2$ plane (left panel of Fig. 3d), and $O_i$ rich regions in the $HgO_y$ layers which correspond to "charge rich" portions in the $CuO_2$ plane (right panel of Fig. 3d).

**Code availability-**The code we used for statistical analysis of the SmXRD data is not currently available (G.C., A.R. and A.B., manuscript in preparation)